\documentclass[11 pt]{article}
\usepackage{graphicx}
\usepackage[ragged]{sidecap}
\usepackage{amsmath, amsfonts, amssymb, mathrsfs}
    \sidecaptionvpos{figure}{c}
    \usepackage[T1]{fontenc}	
\usepackage[utf8]{inputenc}	
\usepackage[english]{babel}	
\usepackage[usenames,dvipsnames]{color}
\usepackage[usenames,dvipsnames,svgnames,table]{xcolor}
\usepackage{lipsum}	
\usepackage{url}		
\usepackage{amsmath, amsfonts, amssymb, mathrsfs}
\usepackage{bm}
\usepackage[english]{varioref}
\usepackage{graphicx}
\usepackage{verbatim}
\usepackage{wrapfig}
\usepackage{subfig}
\usepackage{caption}	
\usepackage[babel]{csquotes}
\usepackage{listings} 
\usepackage{color}
\usepackage[usenames,dvipsnames,svgnames,table]{xcolor}
\usepackage{quoting}
\quotingsetup{font=small}
\usepackage[ragged]{sidecap}
    \sidecaptionvpos{figure}{c}
\usepackage{geometry}
\usepackage{amsmath, amsfonts, amssymb, mathrsfs}
\usepackage{longtable}
\usepackage{multirow}

\def\bicocca{University of Milano-Bicocca, Department of Physics G. Occhialini and INFN, section of Milano-Bicocca}

\def\Title#1{\begin{center} {\Large #1 } \end{center}}
\def\Author#1{\begin{center}{ \sc #1} \end{center}}
\def\Address#1{\begin{center}{ \it #1} \end{center}}

\newenvironment{Abstract}{\begin{quotation}  }{\end{quotation}}
\newenvironment{Presented}{\begin{quotation} \begin{center} 
             PRESENTED AT\end{center}\bigskip 
      \begin{center}\begin{large}}{\end{large}\end{center} \end{quotation}}

\begin{document}
\begin{titlepage}

\vfill
\Title{FLARES}
\vfill
\Author{Mattia Beretta, on behalf of the FLARES Collaboration}
\Address{\bicocca}
\vfill
\begin{Abstract}
FLARES (Flexible Scintillation Light Apparatus for Rare Events Searches) is an innovative project in the field of rare events searches, such as the search for the neutrinoless double beta decay. It aims at demonstrating the high potential of a technique that combines scintillating crystals with arrays of high performance silicon drift detectors (SDD), operated at about 120K, to reach a 2\% level energy resolution in the region of interest ($\sim$ 3 MeV). The proposed technique will combine in a single device all the demanding features needed by an ideal experiment looking for rare events. The characterization of the light emission non-proportionality of different scintillation crystals as well as a first test of a CdWO$_{4}$ crystal coupled to an array of SDD are presented and discussed.
\end{Abstract}
\vfill
\begin{Presented}
 NuPhys2017, Prospects in Neutrino Physics

Barbican Centre, London, UK,  December 20--22, 2017
\end{Presented}
\vfill
\end{titlepage}
\def\thefootnote{\fnsymbol{footnote}}
\setcounter{footnote}{0}

\section{Introduction}

The double beta decay ($\beta\beta$) is a rare spontaneous nuclear transition in which a nucleus (A,Z) decays to a member (A,Z+2) of the same isobaric multiplet. Two final states are possible for this decay: the emission of two electrons and two antineutrinos (2$\nu\beta\beta$) and the emission of two electrons only (0$\nu\beta\beta$). The former process is predicted by the Standard Model of particle physics and has been observed for a dozen of isotopes with half lives ranging from 10$^{18}$ to 10$^{21}$ years. On the other hand, the latter is still object of research, since its discovery would state whether neutrinos are Dirac or Majorana fermions and provide information about their absolute mass scale, while proving the lepton number violation \cite{VisStrum}. These two processes can in principle be distinguished in the energy sum spectrum of the two emitted electrons: while the $0\nu\beta\beta$ results in a peak at the Q-value of the transition (Q$_{\beta\beta}$), $2\nu\beta\beta$ generates a continuous distribution from 0 to Q$_{\beta\beta}$. The observation of the $0\nu\beta\beta$ peak would permit the evaluation of the half life of the decay, directly linked to the effects of the neutrino mass in the process.

\section{Experimental Sensitivity}

In the search for $0\nu\beta\beta$, current generation experiments are struggling to improve their sensitivity, defined as the longest possible half-life measurable with a particular detector above a given background. This characteristic is enhanced on one side by increasing the number of candidate isotopes under observation, the experimental live time and the detector energy resolution, and on the other by reducing the background originating from spurious events in the region of interest (ROI, defined as an energy window equal to one FWHM resolution centered at the Q$_{\beta\beta}$). The sensitivity enhancement requires considerable technical efforts: huge detector masses ($\sim$ton or larger) must be kept efficiently operational for long times, always ensuring low levels of background ($B$) and excellent energy resolution ($\Delta E$). In particular, the background rate of counts is composed by two contributions: the spurious counts due to undesired radioactive sources and the fraction of $2\nu\beta\beta$ events falling in the ROI. While the first component can be reduced with different techniques, for example by shielding the detector and choosing high purity materials, the second is unavoidable and can only be solved with high resolving detectors. The fraction F of $2\nu\beta\beta$ events in the ROI can, in fact, be expressed as \cite{VisStrum}:
\begin{equation} \label{SN2nbb}
F_{2\nu} \backsim Q_{\beta \beta}\cdot \delta^6 = Q_{\beta \beta}\cdot \left(\frac{\Delta E}{Q_{\beta \beta}}\right)^6
\end{equation}
where $\delta$ is defined as the ratio between FWHM resolution and $Q_{\beta \beta}$. This equation shows that resolution plays the most important role in reducing this form of background. At an energy of 3 MeV, a value close to the Q$_{\beta\beta}$ of most of the $0\nu\beta\beta$ candidate isotopes, a $\delta$ of the order of 2\% is needed to make negligible at the actual half life sensitivities of many $0\nu\beta\beta$ experiments the $2\nu\beta\beta$ contribution to the background.


\section{The FLARES strategy}
In this landscape, the FLARES (Flexible Light Apparatus for Rare Events Search) project proposes the development of a detector for $0\nu\beta\beta$ search based on the use of high performance solid state detectors (SDD, Silicon Drift Detectors) to read the light emitted by large scintillating crystals, cooled at $\sim$120K, containing the $0\nu\beta\beta$ candidate isotope. The use of scintillation detectors provide flexibility in the choice of the candidate isotope and easy mass scalability, while the usage of SDDs enhances the attainable energy resolution. Moreover, both scintillating crystals and SDDs can be produced with high purity materials, thus lowering the radioactive contaminations. The project is actually in a research and development phase (R\&D), divided in two parallel aspects: the optimization of the scintillation properties of crystals and the development and characterization of the SDDs.

\subsection{Scintillating Crystals characterization}
\begin{figure}[htb]
\centering
\includegraphics[scale=0.4]{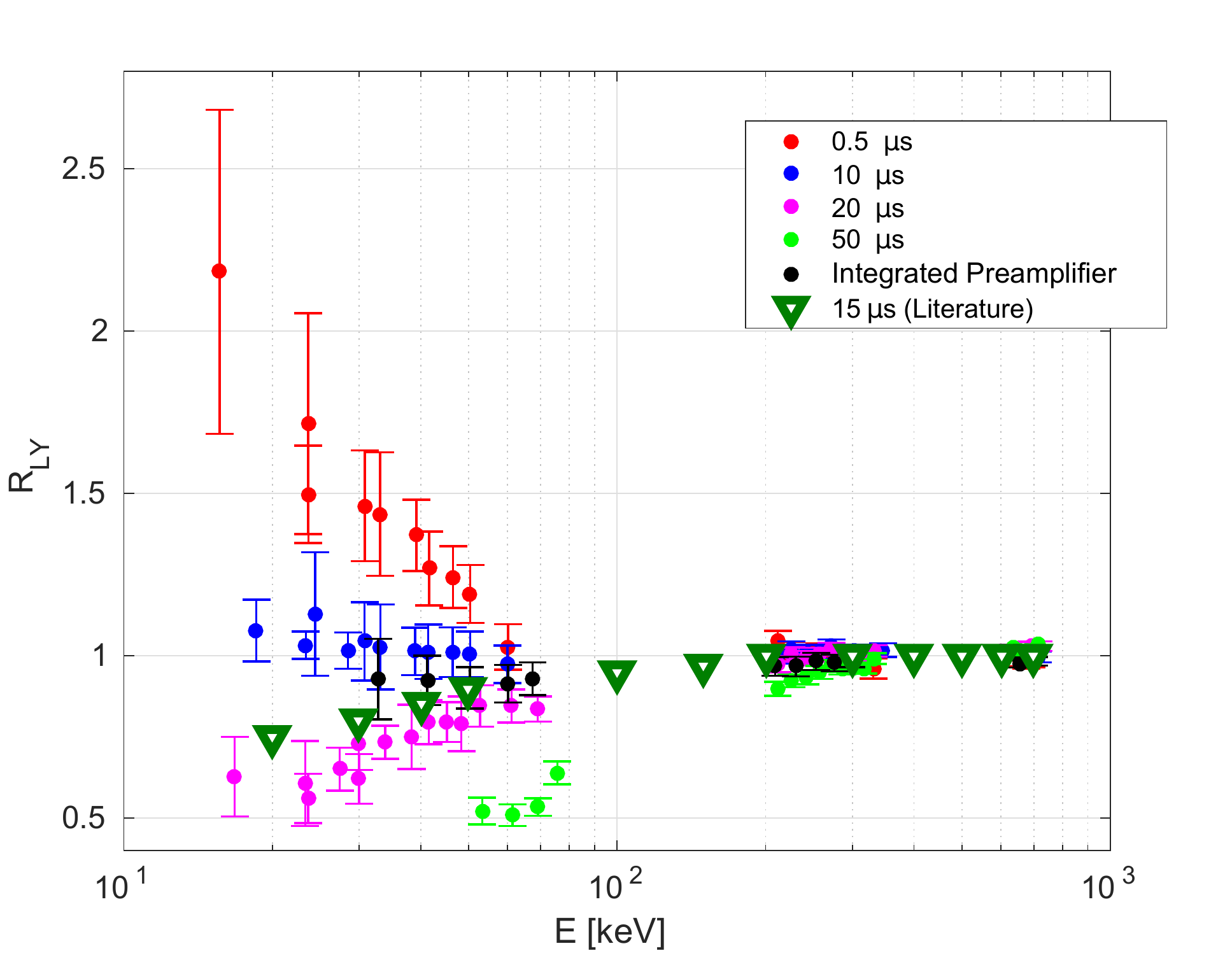}
\caption{Measured non proportionality for CdWO$_{4}$ crystal. The effect of the shaping time is evident,
since it causes the non proportionality to space in a range between 0.5 and 2.5. Integrating the preamplifier signal causes the crystal to reach a minimum value of RLY, therefore of non-proportionality. The figure comes from \cite{FLARES}.}
\label{fig:nonprop}
\end{figure}

\begin{table}[t]
\centering
\begin{tabular}{lllll}  
\hline
	\multirow{2}*{Property}&\multicolumn{2}{c}{CaMoO$_4$}&\multicolumn{2}{c}{CdWO$_4$}\\
						  &	300K	&	120K		& 300K	&	120K\\
	\hline
	Density [g/cm$^3$]	&$\sim$4.3	&	&7.9		&	\\
	Light yield [ph/MeV]	&$\sim$8900	&$\sim$25000		&$\sim$18500	&$\sim$33500	\\
	Scintillation Decay Time [$\mu$s]	&$\sim$18	&$\sim$190	&$\sim$13	&$\sim$22\\
	\hline
	\end{tabular}
	\caption[Properties of CaMoO$_4$ and CdWO$_4$ scintillation crystals]{Properties of CaMoO$_4$ and CdWO$_4$ scintillation crystals. The values come from (\cite{FLARES}).}
	\label{Candidate_Crystals}
\end{table}
One of the first steps to be performed consists in the choice of the scintillating crystal, which has to contain a suitable $0\nu\beta\beta$ candidate isotope while being a performing scintillator. These constraints lead to the selection of two crystals: CdWO$_{4}$ and CaMoO$_{4}$, containing respectively $^{116}$Cd and $^{100}$Mo as candidate isotopes for $0\nu\beta\beta$ decay. The scintillation properties of these two crystals are reported in Table \ref{Candidate_Crystals}. As shown in Table \ref{Candidate_Crystals}, by operating these crystals at a temperature of $\sim$120K, the light yield (LY) increases of a factor 2-3, directly improving the attainable energy resolution. Considering only the statistical fluctuation of information carriers, a resolution of $\leq 1\%$ at 3 MeV can be predicted for a CdWO$_{4}$ crystal operated at 120K \cite{FLARES}. Nevertheless it has to be considered that the attainable resolution for scintillating crystals is usually considered as limited by the intrinsic resolution \cite{MorzRes}. This characteristics depends on the scintillation mechanism of inorganic crystals and could be explained by the lack of proportionality between deposited energies in the lattice and light output, defined as \emph{non-proportionality} of light emission \cite{NonPropComp}. The LY dependency on energy could affect the attainable energy resolution, since the wrong identification of the deposited energy causes the energy peaks of a scintillation spectrum to be wider than expected. As a consequence, the non-proportionality has to be understood in the framework of FLARES project, since it could affect the energy resolution, a key parameter for a rare events detector. A dedicated measurement of the LY non-proportionality of CdWO$_{4}$ has then been carried out, exploiting the Compton Coincidence Technique (CCT)\cite{CCTDes}. Thanks to the CCT, the Relative Light Yield (RLY) has been quantified, defined as the ratio between the light output corresponding to a certain energy deposition and the light output at the 662 keV line of $^{137}$Cs. Values of RLY departing from 1 are due to the non-proportionality, therefore this characteristics has been quantified. To further understand the non-proportionality, the dependence of the RLY on the shaping time (ST) used to filter the scintillation signals has been performed, exploiting a digital trapezoidal shaping algorithm. The results, reported in Figure \ref{fig:nonprop}, show a clear correlation between these two variables \cite{NonPropST}. According to this evidence, the non-proportionality can be modified by changing the ST chosen for the treatment of the pulses. This results implies that an excess with respect to the expected statistical contribution cannot be ascribed only to the crystal inhomogeneities, but has to be traced back to the scintillator/photodetector system. As a consequence, the intrinsic resolution can be reduced and has to be considered as induced by the characteristics of the readout chain and not only by the scintillation mechanism of crystals.

\section{The Silicon Drift Detectors and the first detector test}

The SDDs are solid state detectors with high quantum efficiency ($\sim80\%$ in 450-1000 nm $\lambda$ region) and low electronic noise. In particular, the anode capacintance of these detectors is extremely low ($\sim$ pF) and does not depend on the active area of these devices \cite{SDDGatti}. This advantage of SDDs, as compared to other Si-based photodetectors, allows scaling of detector size to effectively cover larger scintillating crystals with small electronics noise (in particular if the contribution due to leakage current is made negligible by cooling). Consequently, these detectors have been chosen for the FLARES project. In the current R\&D phase, detectors with 9 cm$^2$ active area have been designed by Fondazione Bruno Kessler (FBK). These detectors are the biggest ever designed with a single anode. The first test performed by the means of X-ray measurement showed good noise performances, as well as proving the linearity of their response. 
\begin{figure}[htb]
\centering
\includegraphics[scale=0.4]{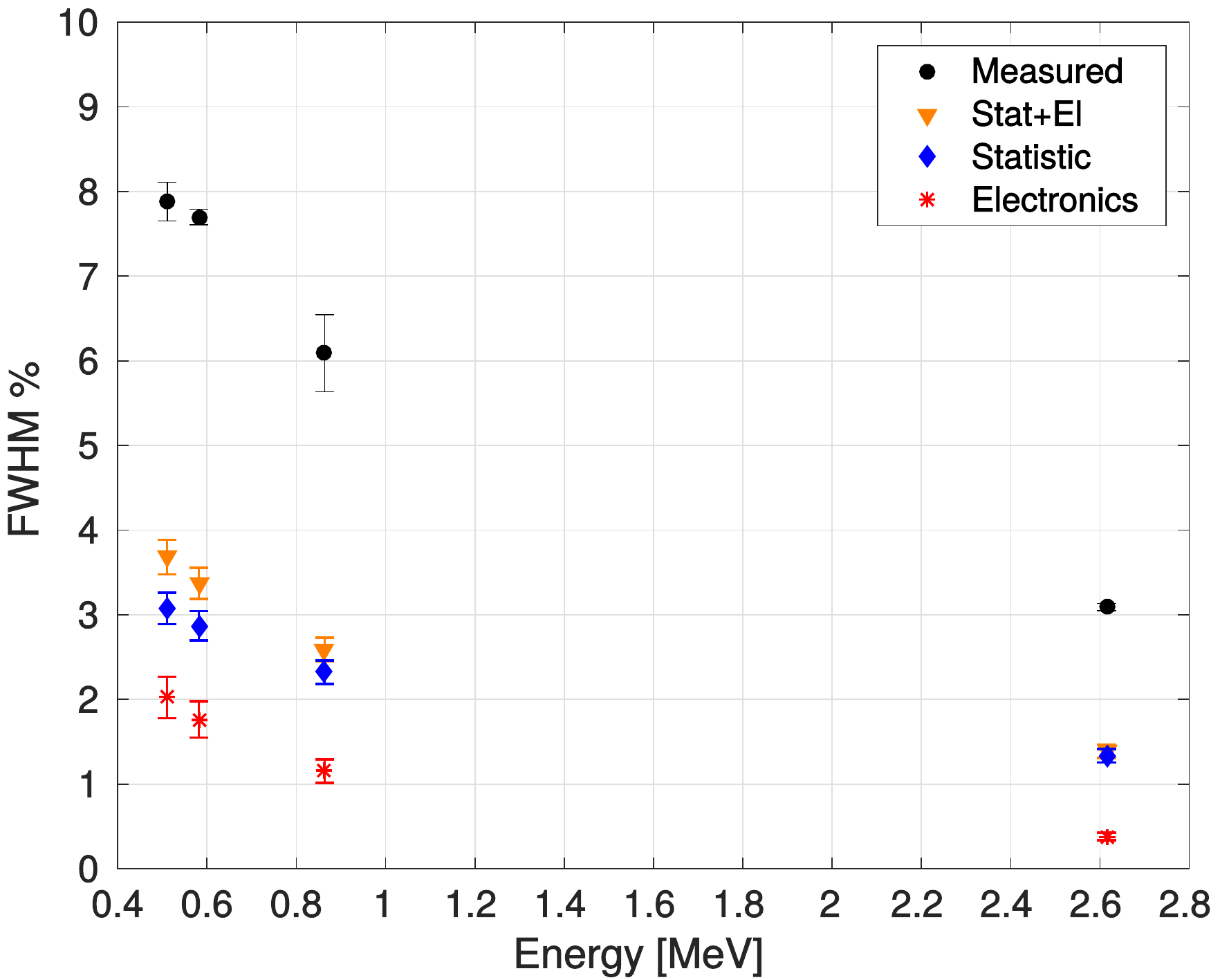}
\caption{Measured FWHM resolution and its components as a function of the $\gamma$-ray energy. The orange points correspond to the squared sum of the statistical (blue points) and electronic (red point) components of the obtained resolution. The figure comes from \cite{SDDCdWO4}.}
\label{fig:res}
\end{figure}

\noindent
Furthermore, a measurement with a CdWO$_4$ coupled to an SDD matrix has been performed in collaboration with the Politecnico of Milan, department of Electronics, Information and Bioengineering. The detailed results have been recently published in \cite{SDDCdWO4}. With a $^{228}$Th source measurement, a 3\% FWHM energy resolution has been obtained for the 2615 keV $\gamma$ peak (see figure \ref{fig:res}). The measured resolution appears limited by an exceeding component, that could be ascribed to photoelectron losses when the photon interaction happens far from the SDD center. This results shows the effectiveness of FLARES strategy to resolving scintillators, thus preliminary proving this detector concept. Further measurements with more optimized SDDs are currently being performed, with the objective of reaching resolution values suitable for the application in the search of rare events.

\end{document}